\documentclass[fleqn,usenatbib]{mnras}

\usepackage{newtxtext,newtxmath}
\usepackage[T1]{fontenc}

\DeclareRobustCommand{\VAN}[3]{#2}
\let\VANthebibliography\thebibliography
\def\thebibliography{\DeclareRobustCommand{\VAN}[3]{##3}\VANthebibliography}

\usepackage{graphicx}	% Including figure files
\usepackage{amsmath}	% Advanced maths commands
\title[FRB relativistic motion]{Constraint on the relativistic motion of fast radio bursts based on the maximal electric field}

\author[Shen \& Zou]{
Jun-Yi Shen,$^{1}$
and Yuan-Chuan Zou$^{1}$\thanks{E-mail: zouyc@hust.edu.cn}
\\
$^{1}$ Department of Astronomy, School of Physics, Huazhong University of Science and Technology, Wuhan 430074, China
}

\date{Accepted XXX. Received YYY; in original form ZZZ}

\pubyear{2023}

\begin{document}
\label{firstpage}
\pagerange{\pageref{firstpage}--\pageref{lastpage}}
\maketitle

\begin{abstract}
Fast radio bursts (FRBs) are millisecond radio signals from cosmological distances. As they propagate, FRBs can interact with ambient photons and initiate a quantum cascade that can limit the electric field strength. This paper examines whether some observed bright and brief FRBs may challenge this limit if the source is not relativistic. The size of a static FRB source is estimated as $R\sim ct$, where $t$ is the time scale of the FRB and $c$ denotes the speed of light. But for a relativistic source moving at the Lorentz factor $\Gamma$, the size is $R \sim 2 \Gamma^2 c t $. Using an FRB catalog, we plot the luminosity-duration distribution. Most FRBs fall below the limit for a static source, but two events have higher luminosity and shorter duration. This suggests these bursts may originate from relativistic sources, although more data is needed to confirm this. %Further study of more extreme FRBs may constrain the fraction that require relativistic beaming.
\end{abstract}

\begin{keywords} transients: fast radio bursts -- plasmas   

\end{keywords}

%%%%%%%%%%%%%%%%%%%%%%%%%%%%%%%%%%%%%%%%%%%%%%%%%%

%%%%%%%%%%%%%%%%% BODY OF PAPER %%%%%%%%%%%%%%%%%%

\section{Introduction}
Fast radio bursts (FRBs) were first discovered by Lorimer in 2007 as millisecond radio transient sources \citep{2007Sci...318..777L}. 
They are radio signals that originate from cosmological distances. 
The luminosity of individual FRB event ranges from $10^{38}$ erg $\mathrm{s^{-1}}  $\citep{2020Natur.587...59B} to $ 10^{46}$ erg $\mathrm{s^{-1}}  $\citep{2019Natur.572..352R}. 
The origin of FRBs is still mysterious. 
For non-repeating events, some catastrophic events have been proposed, such as mergers of binary white dwarfs \citep{2013ApJ...776L..39K}. 
Observations by the Canadian Hydrogen Intensity Mapping Experiment (CHIME) collaboration suggest that at least a portion of FRBs is associated with magnetars \citep{2020Natur.587...59B}. 
FRBs could be generated by magnetar's charge starvation region and current sheet region \citep{2022ApJ...925...53Z}. 
More recent review on the physics of FRBs can be seen in \cite{2022arXiv221203972Z}.

The brightness temperature of FRB is exceptionally high \citep{2022arXiv221203972Z}: $T_{{\rm{B}}} \simeq 10^{36} \  \rm{K}$.
This indicates that the FRBs must be generated by a coherent process, while the exact radiation mechanism and origin are still unknown. 
Several models have been proposed to address this issue, as discussed in the review by \citet{2020Natur.587...45Z}. 
Some models propose a  {gamma-ray burst-like mechanism (GRB-like), that is,} FRBs are produced by relativistic shocks, as suggested by \cite{2014MNRAS.442L...9L}. Some models utilize a pulsar-like model, where coherent bunches of radio signals are emitted as FRBs, as described by \cite{2000ApJ...544.1081M}. 
%\cite{2022ApJ...925...53Z}  investigates one approach to generating FRBs through coherent bunches.  
One difference between the GRB-like and pulsar-like models is that the GRB-like models require a relativistic jet radiation region, while this is not necessary for the pulsar-like models. 
It would be meaningful to investigate whether the source of FRBs is relativistic or static.

Studying extreme events characterized by high luminosity and short duration can provide insights into the source's nature. 
%The FRB 20150807A and FRB 20191108A are the examples \citep{2020TNSFR2470....1P}. 
 {One common estimate of the size of an astrophysical source is $R \sim c t $, with $t$ being the time scale and $c$ being the speed of light.
Extreme events have high luminosity and a small source size $R$,} resulting in a high energy density.  
However, there are limits to the energy density. 
One of them is the strength of the corresponding electric field of the electromagnetic wave $ E < E_{{\rm{s}}} \equiv m_{{\rm{e}}}^2 c^3 / \hbar e = 1.32 \times 10^{16} $ V $\mathrm{cm^{-1}}$, where $m_{{\rm{e}}}$ is the mass of the electron and $e$ is the charge of the electron and $\hbar$ is the Planck constant. 
Once if $E$ reaches this limit, the energy of electromagnetic field converts into kinetic and rest-mass energy of electron-position pairs \citep{PhysRev.82.664}. \citet{2022ApJ...929..164Z} found a stronger constraint on $E$ based on their simulations, which is $E< 3 \times 10^{12} \ {\rm V\,cm^{-1}}$. 
With this limit they also placed constraints on the size of the radiation site. 
With the known duration of the FRBs, we may also constrain the Lorentz factor of the emitting region if the source is relativistic.

%The luminosity of an FRB can be inferred by its dispersion measure (DM) and flux, which are obtained through observations. DM is a measure of the effect of radio signal propagation, where the velocity of electromagnetic waves in plasma depends on their frequency. In the case of FRBs, different frequency components experience time delays. By utilizing DM, the luminosity distance can be estimated, enabling further cosmological studies such as the Hubble tension  \citep{2022MNRAS.515L...1W} and upper limit of photon mass \citep{2016PhLB..757..548B}.
 
%The GRB-like model postulates that the source region of FRBs has a significant bulk Lorentz factor, $\Gamma$ \citep{2014MNRAS.442L...9L}. By constraining the Lorentz factor $\Gamma$ of the source, we can evaluate the validity of the GRB-like model. 
%For FRBs to propagate through a plasma, the frequency of the electromagnetic wave must be higher than the plasma frequency. This allows us to impose constraints on the $E$, further constraints on the $\Gamma$.

This paper is organized as follows. Section \ref{sec2} introduces the quantum cascade process of the interaction between FRBs and high-energy photons, demonstrating the upper limit of $E$. Section \ref{sec3} highlights the distinctions between a relativistic source and a non-relativistic source. In section \ref{sec:data}, we present the selected data to examine the validity of the relativistic source hypothesis. Finally, in section \ref{sec5} we draw our conclusions based on the findings presented in this paper.

\section{Interaction between FRB and photons}\label{sec2}
From the observations, we can estimate the electric field strength of the FRBs to be greater than $10^{9} \ \rm{V \, cm^{-1}}$ \citep{2022ApJ...929..164Z}. 
In such a strong field, the interaction between high-energy photons and the field has been extensively studied \citep{Kirk_2009}, including processes such as the Breit-Wheeler process and nonlinear inverse Compton scattering, which are quantum-electrodynamics (QED) processes. The Breit-Wheeler process involves the interaction of two high-energy photons, resulting in the creation of an electron-positron pair. The FRB GHz photons can be upscattered by high-energy electrons through inverse Compton scattering. 
If this scattering process generates a significant number of high-energy photons, providing sufficient energy for subsequent pair creation, the cascade starts. The presence of high-energy photons acts as a trigger, initiating the reaction.
Then the Breit-Wheeler process occurs and creates positron-electron pairs. The electrons and positrons accelerate in electromagnetic field and radiate high energy photons. These high-energy photons continue to create pairs. The reaction rate is (in weak field) \citep{landau1982quantum}:
\begin{equation}
    p=\frac{ 3^{3/2} |e|^3 E }{ 2^{9/2} m_{{\rm{e}}} }\mathrm{exp(-\frac{8}{3\eta}}),
    \label{eq2}
\end{equation}{}
with
\begin{equation}
    \eta=\frac{\hbar^2 |e| \gamma \omega }{ m_{{\rm{e}}}^3 c^5 }\sqrt{(\mathbf{E} + \mathbf{v} \times \mathbf{B})^2 - (\mathbf{v} \cdot \mathbf{E})^2/c^2},
    \label{eq3}
\end{equation}
where $\omega$ is the angular frequency of electromagnetic field, and the $\gamma$ is the Lorentz factor of electrons.
There is a critical electric field $E_s$, which is equal to $ 1.32 \times 10^{16} $ V $\mathrm{cm^{-1}}$. 
When the field strength $E$ of an electromagnetic wave reaches this critical value, the cascade process is expected to occur rapidly.  
This leads to a quick decrease in the energy of the electromagnetic field. 
\cite{2019MNRAS.483L..93L} investigated that this upper limit of $E$ is a constraint on the maximum luminosity of FRB. 
Typically, when $\eta \sim 5 \times 10^{-2}$, the quantum cascade takes place. From Eq. (\ref{eq3}), it can be inferred that higher field intensity $E$ leads to higher reaction rate, resulting in a rapid pair cascade and the formation of a dense plasma. 

The cascade process causes the transfer of energy from the electromagnetic field to the plasma, resulting in an increased background plasma density and a decrease in the electromagnetic field strength. To ensure the propagation of electromagnetic waves in the medium, the condition $\omega > \omega_{\rm{p}}$ must be satisfied for electromagnetic waves to propagate, where the $\omega_{{\rm{p}}}$ is the plasma frequency. 
We have: 
\begin{equation}
    \omega_{{\rm{p}}} = \sqrt{\frac{4 \pi n  e^2}{  m_{{\rm{e}}}}},
\end{equation}
where $n$ represents the plasma density, which must be sufficiently low for electromagnetic waves to propagate. However, the cascade process increases plasma density $n$. Therefore, it imposes a constraint on the electric field $E$ of the electromagnetic wave. 

\cite{2022ApJ...929..164Z} used simulations to determine that the electric field $E$ of an FRB must be smaller than $3 \times 10^{12} $ V $\mathrm{cm^{-1}}$ (in the laboratory reference frame), ensuring that the FRBs can propagate in the medium. This new constraint is much smaller than the $E_s$. There should be a new constraint on the physical properties, such as the radiation radius, the Lorentz factor of the emitting region, and the maximum luminosity. \cite{2022ApJ...929..164Z} has already placed a lower limit of the radiation radius. In this work, we focus on the Lorentz factor.

\section{the state of source } \label{sec3}
In some models, it is predicted that FRB sources exhibit GRB-like behavior and must be in relativistic motion, with a Lorentz factor $\Gamma \sim 10^2$ \citep{2014MNRAS.442L...9L}. These are the so-called GRB-like models, which radiate FRB from relativistic jet. In this section, we aim to explore the disparities between static and relativistic sources with a focus on their geometries. Notably, highly relativistic motion leads to the beaming effect. Considering FRBs as plane waves, we denote the energy flux density as: 
% The scale of the source is $R = c \delta t$.
$S_{0} = E^2 c / 4\pi $ (in cgs units). 
Therefore, for an isotropic static source, the FRB luminosity is:
 \begin{equation}
     L=E^2 c R^ 2.
     \label{L5}
 \end{equation} 
 % We can plot this relationship in log figure, see figure.1 black solid line. The intercept denotes the E. If the FRB simple reach the black solid line, which means the E larger than the constraint $E_{max} = 3 \times 10^{12}$ V $\mathrm{cm^{-1}}$ (lab reference), the complain is the static source assumption is incorrect. The lorenz factor is a considerable value.
For the static and relativistic sources, the Eq. (\ref{L5}) has different forms (see the discussion below). 

%\subsection{\textcolor{red}{static source and relativistic source} }
Assuming the source is isotropic and the radiation is simultaneous, the duration of FRB $t$ is the result of the time delay during propagation. The scale of the source is:
\begin{equation}
    R \sim c t.
    \label{eq6}
\end{equation}
Due to the differences in geometry, the scale of the source in a static configuration differs from that of a relativistic source.

In the case of relativistic sources, the radiation is concentrated within a cone of angle  $ 1/\Gamma $,  {because of the relativistic beaming effect.} %( { We assume the jet angle is greater than $1/\Gamma$, meaning that only radiation within a special cone is detectable, but not the whole jet. Within this limited region, the luminosity exhibits tiny fluctuation. Therefore, We use isotropic luminosity of FRBs to calculate. }) 
Therefore, in areas where the angle between the line of sight and the line connecting the observer's position and the source center is larger than $ 1/\Gamma $, the radiation will not propagate to observer. 
Consequently, not all parts of the source are observable. This geometry differs from that of static sources. 
In the observer's line of sight, the source appears as a cone. 
In this scenario, the radius and time $t$ obeys:
\begin{equation} \label{eq7}
    R \sim 2 \Gamma^2 c   t .
\end{equation}

The difference between relativistic sources and non-relativistic sources lies in its geometry, as shown in Eq. (\ref{eq6}) and Eq. (\ref{eq7}).

% \begin{figure}
% 	% To include a figure from a file named example.*
% 	% Allowable file formats are eps or ps if compiling using latex
% 	% or pdf, png, jpg if compiling using pdflatex
% 	\includegraphics[width=\columnwidth]{f1.png}
%     \caption{Schematic diagram for the non-relativistic source. }
%     \label{fig:1}
% \end{figure}
% \begin{figure}
% 	% To include a figure from a file named example.*
% 	% Allowable file formats are eps or ps if compiling using latex
% 	% or pdf, png, jpg if compiling using pdflatex
% 	\includegraphics[width=\columnwidth]{f2.png}
%     \caption{Schematic diagram for the relativistic source. Emission is beamed in a narrow angle $1/\gamma$.}
%     \label{fig:2}
% \end{figure}

\section{Constraints from the FRB data } \label{sec:data}
In Section 3, it is described that by imposing constraints on $\Gamma$, it is possible to determine whether $\Gamma$ of the GRB-like model can be confirmed.  Applying Eq. (\ref{eq6}) to Eq. (\ref{L5}), we have 
\begin{equation}
    L = E^2 c^3   t^2.
    \label{lnr}
\end{equation}
Eq. (\ref{lnr}) shows the radiation luminosity of a static source and can be plotted on a log-log scale. If the static assumption is not valid (breakthrough the $E_{\max} \equiv 3 \times 10^{12}\rm{V\;cm^{-1}}$), the Eq. (\ref{lnr}) should be swapped to ( {isotropic equivalent luminosity}) \footnote{ {Here we assume the relativistic jet opening angle is greater than $1/\Gamma$. Because of the beaming effect, the main emission is from the region inside $1/\Gamma$. Therefore, we do not need a jet angle correction  for the observed the luminosity.}}:
\begin{equation}
    L= 4 E^2  \Gamma^4 c^3   t^2.
\end{equation}
with the same maximum $E$, this relativistic source could supply higher luminous events. 
The difference can be shown in a  {diagram $L-t$}, which will be shown later.
%Our propose is to use the observational data to plot this figure. The goal is to obtain the Lorentz factor from the $L-t$ (luminosity-duration) diagram. In this section we will show our figure.

\subsection{DM of FRBs} \label{sec:DM}
DM describes the effect of propagation for electromagnetic waves, particularly in the case of cosmological distance events such as FRBs.  In the range of FRBs band, the different wavelengths propagate in the plasma of IGM with varying velocities. This can be described by the DM:
\begin{equation}
    \mathrm{DM}=\int \frac{n_{{{\rm{e}}},z}}{1+z} \mathrm{d} l,
\end{equation}
where $\rm{n_{e,z}}$ is the particle number density at redshift $z$, and $\mathrm{d} l$ is an infinitesimal proper length along the line of sight.
DM has three parts, which can be written as :
\begin{equation}
    \rm {DM_{obs}=DM_{MW}+DM_{IGM}+DM_{host} }.
\end{equation}
The term $\rm{DM_{MW}}$ denotes the Milky Way contribution to DM, which can be estimated by model NE2001 \citep{2002astro.ph..7156C} and YMW2016 \citep{2017ApJ...835...29Y}.  {NE2001 and YMW2016 are models that describe the electron distribution in the Milky Way. Once the cardinal direction of the FRB (gl gb) has been cleared, these two models can provide an estimate of $\rm{DM_{MW}}$ for one event.}
The term $\rm{DM_{host}}$ represents the contribution to DM from the host galaxy, which is not well studied and is often assumed to follow a log-normal distribution \citep{2020Natur.581..391M}.
 The $\rm{DM_{IGM}}$ component accounts for the dispersion measure contributed by the intergalactic medium (IGM), which contains information about the IGM over long distances. 
 $\rm{DM_{IGM}}$ can be used to infer the redshift ($z$) of the FRB event. The relation between $\rm{DM_{IGM}}$ and $z$ is given by \citep{2022arXiv221203972Z}:
 \begin{equation}
     \mathrm{\langle  DM_{IGM} \rangle}=\frac{3 c H_0 \Omega_b f_{\mathrm{IGM}}}{8 \pi G m_p} \int^z_0 \frac{\chi(z) (1+z) dz}{[\Omega_m (1+z)^3 + \Omega_{\Lambda}]^{1/2}},
     \label{eq12}
 \end{equation}
%介绍一下这里式子的含义xxx
where $\Omega_b=0.0486, \Omega_m=0.3089, \Omega_\Lambda=0.6911$ are cosmology constants, chosen from the PLANCK observation \citep{2016A&A...594A..13P}, and $H_0$, $f_{\rm{IGM}}$, $G$, $m_{{\rm{p}}}$  
$\chi(z)$ denotes  Hubble constant,  the fraction of baryon mass in the IGM, gravitational constant, the proton mass, the ratio of the number
of free electrons to baryons in the IGM, respectively. 
Due to fluctuations in the IGM, we can only obtain the expectation value of $\rm{DM_{IGM}}$. The fluctuation of $\rm{DM_{IGM}}$ can be expressed as \citep{PhysRevD.100.083533}: 
\begin{equation}
   \mathrm{ \frac{\sigma_{\Delta IGM}}{\langle  DM_{IGM} \rangle}}=\frac{0.2}{\sqrt{z}}.
   \label{eq13}
\end{equation}

\subsection{Luminosity -- duration diagram of FRBs}\label{sec:L-t}
\begin{figure}
	\includegraphics[width=\columnwidth]{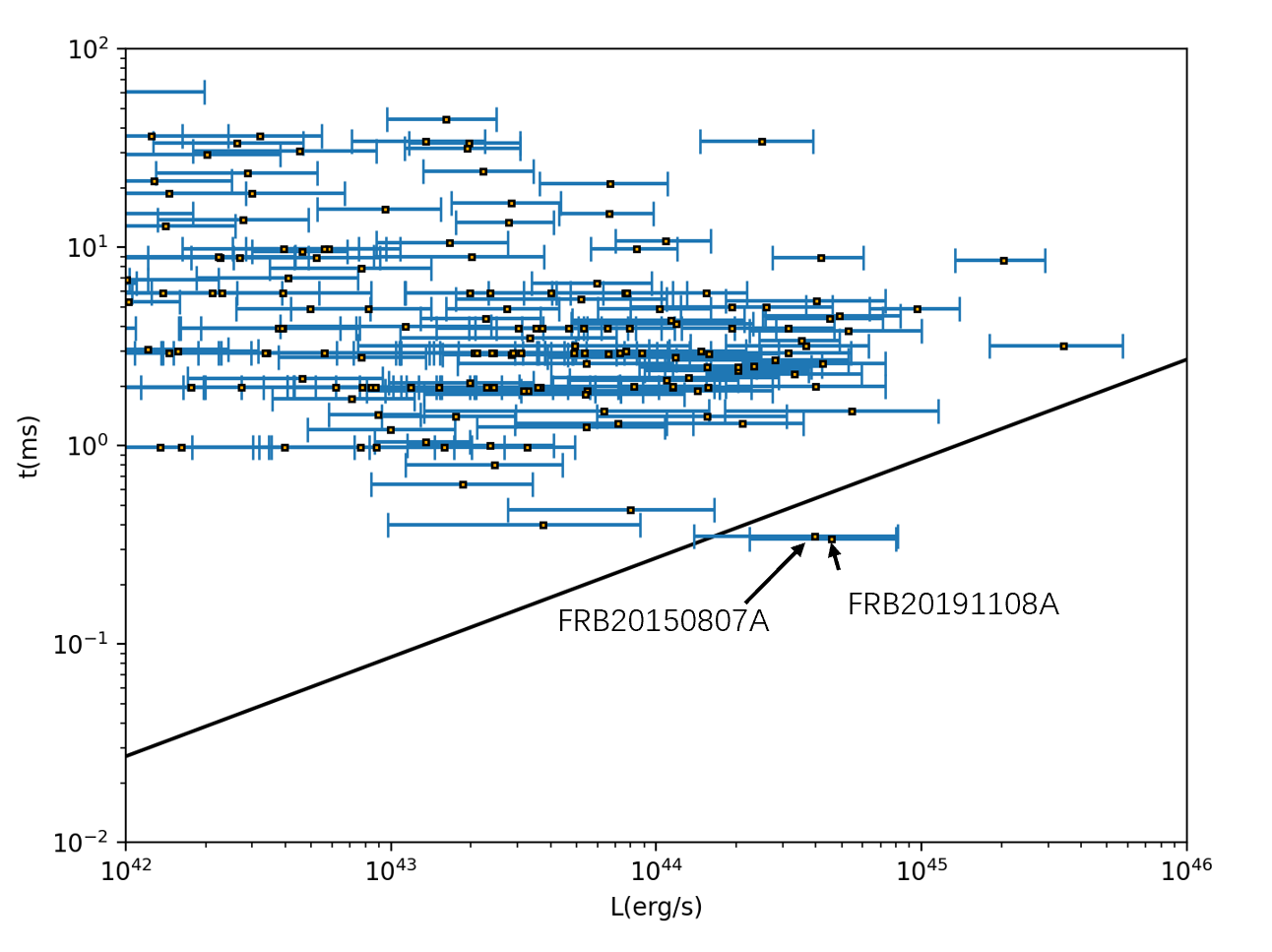}
    \caption{The $L$ (luminosity)- $t$ (duration) scattered plot of FRBs. Luminosity is calculated from Eq. (\ref{L}) with $\nu=\nu_c$. The $t$ is full width at half maximum of the pulse. The black solid line is plotted following the Eq. (\ref{lnr}) at ${E=E_{\max}}$.  Two FRBs (FRB20150807A and FRB20191108A) of them locate at the more luminous side. }
    \label{fig:fig3}
\end{figure}

After obtaining $\rm{DM_{IGM}}$ and inferring the redshift $z$ using Eq. (\ref{eq12}), we can calculate the uncertainty of $\rm{DM_{IGM}}$ using Eq. (\ref{eq13}). This enables us to determine the uncertainty in the $z$ \footnote{$F_{\nu,p}$ also will affect the uncertainty of luminosity $L$. However, $F_{\nu,p}$ is detected precisely. The main uncertainty comes from the detection of z.}. The luminosity distance $D_{{\rm{L}}}$ can be inferred from $z$:
\begin{equation}
    L=4 \pi D_{{{\rm{L}}}}^2 F_{\nu,p} \nu, 
    \label{L}
\end{equation}
where $D_{{\rm{L}}}$ is the luminosity distance, which can be determined from the redshift $z$, and the $F_{\nu,p}$ is the peak flux density. Since observations cannot cover entire band of an FRB, we need to estimate the $\nu$. One common estimate is to take the center frequency $\nu_c$ as a rough approximation. Another option is to use the telescope bandwidth $\nu_o$ as a conservation estimate. $D_{{\rm{L}}}$ can be firmed by redshift $z$, and the $F_{\nu,p}, \nu_c$ are observational data. 
We use the data from the FRBSTATS Catalogue\footnote{\href{https://www.herta-experiment.org/frbstats/catalogue}{https://www.herta-experiment.org/frbstats/catalogue}},
which are from CHIME, GBT, FAST, ASKAP, and others \citep{2016PASA...33...45P,2020TNSFR2470....1P,2021ascl.soft06028S}. 
The data provided in the catalogue include peak flux density, inferred redshift $z$, bandwidth, and center frequency $\nu_c$. 
For repeating bursts, only the highest signal-to-noise event is included. There are 828 events (source) in total in this dataset. Redshift $z$  is induced by DM as described in Section \ref{sec:DM}. 
The data is updated to December 2022, and the latest event is FRB20221128.
 {The scattered plot of $L$ (luminosity) - $t$ (duration) is shown in Fig. \ref{fig:fig3}.} 
We only choose those FRBs with luminosity larger than $ 10^{45}$ erg $\rm{s^{-1}}$ as the even dimmer FRBs are unlikely to break the limit line as given in Eq. (\ref{lnr}), taking the $E_{\max}=3 \times 10^{12} \,\rm{V\,cm^{-1}}$ into this equation.
This limitation line is shown as a straight line in Fig. \ref{fig:fig3}.
For a certain FRB locating on the right side of this line, it indicates at least this FRB should not radiate stationarily.
Since \cite{2022ApJ...929..164Z} only simulated  {the electromagnetic wave interaction of the 1 GHz FRB}, we only chose these FRBs with frequencies in the range of $0.6$-2 GHz.

Based on Fig. \ref{fig:fig3},  it can be seen that the majority of the FRB events are compatible with the stationary scenario.
However, there are two events FRB20150807A (duration $t$=0.35 ms, redshift $z$=0.27, and peak flux $F_{\nu,p}$=128 Jy ms) and FRB20191108A  (duration $t$=0.34ms, redshift $z$=0.56, and peak flux $F_{\nu,p}$=27 Jy ms) \citep{2020TNSFR2470....1P} that transcend the limit of the constraints, indicating that they may need to be relativistic. 
Considering that they are not clearly far from the limitation line, we cannot conclude that FRBs should radiate in relativistic velocity, and cannot constrain the Lorentz factor neither. 
%This observation suggests that the sources of these FRB events are constant with the  non-relativistic motion assumption. The outcome also shows that FRBs sources do not have to be in relativistic motion. Most of FRBs source in non-relativistic motion obey the constraints on $E$ taken out in this article.

\section{conclusion and discussion} \label{sec5}
In this work, we examined whether some bright, short duration FRBs could originate from relativistic sources, based on a quantum cascade limit on the electric field strength. 
The luminosity-duration distribution for 828 FRB events were analyzed. Most FRBs fell below the limit expected for a static, non-relativistic source. 
 {However, the two bursts had a higher luminosity and a shorter duration,} suggesting they may originate from relativistic sources. 
This could support GRB-like radiation models for part of the FRBs. 
However, more extreme FRBs are needed to strengthen this conclusion.

%The analysis presented provides initial hints that a fraction of FRBs could arise from relativistic, beamed sources. This would point to explosive, catastrophic origins like GRBs or magnetar flares, rather than pulsations from rotating neutron stars. However, the majority of known FRBs satisfy limits for static emission, so likely have a mix of progenitors. More extreme FRBs pushes past the quantum cascade limit will help identify the fraction that require relativistic motion.

If confirmed, the Lorentz factor distribution could distinguish GRB-like and pulsar-like models. The progenitor environments and emission physics may also differ between relativistic and non-relativistic FRB sources. 
For the GRB-like model, it predicts multiwavelength afterglow.
Further multi-wavelength follow-up will shed light on the origin. 
The radio spectra and polarization properties could also diagnose relativistic boosting effects. Ultimately, identifying more high luminosity, short duration FRBs and modeling their radiation mechanisms will help unravel the origin of these enigmatic signals.

\section*{Acknowledgements}
We thank the helpful discussions with Enping Zhou,  Hao Wang, Shiyan Tian, Lin Zhou, Yuanyuan Zuo and Ruihang Dong. The English is polished with ChatGPT.
This work is supported by the National Natural Science Foundation of China (Grant No. 12041306).

%%%%%%%%%%%%%%%%%%%%%%%%%%%%%%%%%%%%%%%%%%%%%%%%%%

\section*{Data Availability}
The data used are publicly available  {in the FRBSTATS catalog} (\url{ https://www.herta-experiment.org/frbstats/catalogue}).



\bibliographystyle{mnras}
\bibliography{FRB-Gamma} % if your bibtex file is called example.bib

%\begin{thebibliography}{99}
%\end{thebibliography}

%%%%%%%%%%%%%%%%%%%%%%%%%%%%%%%%%%%%%%%%%%%%%%%%%%

%%%%%%%%%%%%%%%%% APPENDICES %%%%%%%%%%%%%%%%%%%%%

%%%%%%%%%%%%%%%%%%%%%%%%%%%%%%%%%%%%%%%%%%%%%%%%%%

% Don't change these lines
\bsp	% typesetting comment
\label{lastpage}
\end{document}